# Understanding Controls on Interfacial Wetting at Epitaxial Graphene: Experiment and Theory

Hua Zhou[1*], P. Ganesh[2†], Volker Presser[3], Matthew C. F. Wander[4], Paul Fenter[1‡], Paul R. C. Kent[2], De-en Jiang[5], Ariel A. Chialvo[5], John McDonough[3], Kevin L. Shuford[4] and Yury Gogotsi[3]

[1] Chemical Science and Engineering Division, Argonne National Laboratory, Argonne, IL 60439

[2] Center for Nanophase Materials Sciences, Oak Ridge National Laboratory, Oak Ridge, TN 37831.

[3] Department of Materials Science and Engineering & A.J. Drexel Nanotechnology Institute, Drexel University, Philadelphia, PA 19104

[4] Department of Chemistry, Drexel University, Philadelphia, PA 19104

[5] Chemical Science Division, Oak Ridge National Laboratory, Oak Ridge, TN 37831

## Abstract

The interaction of interfacial water with graphitic carbon at the atomic scale is studied as a function of the hydrophobicity of epitaxial graphene. High resolution X-ray reflectivity shows that the graphene-water contact angle is controlled by the average graphene thickness, due to the fraction of the film surface expressed as the epitaxial buffer layer whose contact angle (contact angle $\theta_c$ = 73°) is substantially smaller than that of multilayer graphene ($\theta_c$ = 93°). Classical and *ab initio* molecular dynamics simulations show that the reduced contact angle of the buffer layer is due to both its epitaxy with the SiC substrate and the presence of interfacial defects. This insight clarifies the relationship between interfacial water structure and hydrophobicity, in general, and suggests new routes to control interface properties of epitaxial graphene.

Work performed under the auspices of the Office of Science, Division of Chemical Science, US-DOE under contract No. DE-AC02-06CH11357

# I. INTRODUCTION

An understanding of the fundamental interactions at liquid-graphitic carbon interfaces is essential to their use in electrochemical energy storage systems, including batteries (redox reactions) and supercapacitors (non-Faradic reactions) [1, 2]. Significant efforts have investigated the interaction of graphitic substrates with liquids to identify the driving forces that mediate these interactions [3-7]. In particular, it has been observed that graphene – water interaction (as probed by its contact angle) can be systematically controlled, although this phenomenon is not well-understood [8]. High-resolution experimental and computational validation of these observations is absent. Although there is great potential for using graphene and graphene-based materials in energy storage and conversion systems [9, 10], to date the most detailed studies concern the interaction of graphitic carbon with adsorbed films of crystalline ice instead of liquid water [11, 12]. This is partly because it is experimentally challenging to probe and quantify the atomic structure of water-carbon interfaces, especially for highly textured graphitic carbons (such as crumpled graphene sheets, nanotube bundles, and onion-like carbon powders) [13-15].

Recent advances in synthesizing atomically flat two-dimensional carbon sheets, such as free-standing graphene [16] and "epitaxial graphene" (EG) grown on single-crystal silicon carbide (SiC) [17], open a new route to understand water-carbon interactions at ambient conditions with a truly atomic-scale perspective. Accurately describing the interfacial structure and dynamics between water and graphene is a first step towards understanding more complex fluid-solid interfaces (*e.g.*, organic electrolytes and ionic liquids).

The present work derives from our unexpected observation that the macroscopic contact angle of water on EG films depends on the EG film thickness (i.e., from zero-layer to multi-layer graphene). To understand this observation, we integrated high resolution X-ray reflectivity (XR) measurements with computational approaches, including molecular dynamics (MD) simulations and density functional theory (DFT) calculations. Our study shows that the epitaxial buffer layer ($G_0$) is more hydrophilic than subsequent layers ($G_n$) owing to the increased perturbation of $G_0$ by the SiC substrate, surface defects and functional groups. Together, experiments and simulations provide a fully molecular-scale understanding of the interfacial water structure and the structural changes associated with changes in wetting properties and confirm a strong correlation between interfacial bonding, hydrophobicity, and the macroscopic contact angle. These observations shed new light on electrochemistry [18] and functionalization [19] of graphene, and bear significant implications for the understanding and control of hydrophobic interactions in fields as diverse as protein folding and self-assembly [20].

## II. METHODS

### A. Experimental methods

#### a. Epitaxial graphene growth and characterization:

Epitaxial graphene (EG) film was grown on on-axis cut 6H-SiC (0001) wafer (nitrogen doped; Cree Inc., USA) via thermal decomposition in a vacuum oven (Solar Atmospheres, PA). Vacuum decomposition, as illustrated in Fig. 1(A), was carried out at 1200 – 1500 °C in a high vacuum of $10^{-6}$ Torr (heating rate: 10 K/min). In this work, a set of EG samples with various film thickness and layer coverage were grown by tuning the annealing temperature and time.

EG samples were characterized by both Raman Spectroscopy and Atomic Force Microscopy (AFM). Raman spectra were recorded with an inVia Renishaw (Gloucestershire, UK) microRaman-spectrometer. An Ar-ion laser with horizontal polarization was operated at 514.5 nm in a backscattering geometry. The spectral resolution was 1.7cm$^{-1}$ (1800 lines/mm grating) and the lateral resolution was 0.7 μm. Fig. 1 (B-D) shows Raman spectra of graphene and the underlying SiC substrate. The spectra from a bilayer graphene (BLG) sample (see below) are consistent with previous work [21]. The full width at half maximum (FWHM) of the 2D band is in a good agreement with that predicted for BLG based on the expected inverse linear relationship between FWHM and the number of layers [21]. Subsequent X-ray reflectivity analysis corroborated this layer thickness (see below). Surface morphology of EG was imaged in Tapping Mode using a MultiMode, Digital Instruments Nanoscope IIIa AFM. The measurements were performed in air at room temperature. Both height and derivative images were acquired simultaneously from scans across the surface of a few-layer EG sample [Fig. 1(E) and (F)]. The step-terrace morphology inherited from the SiC substrate is clearly seen. The derivative image enhances the contrast between terraces and adjacent graphene layers. The surface topography obtained is consistent with the structural features of graphene layers resolved from X-ray reflectivity analysis.

The wetting properties of a series of EG films were also characterized by contact angle measurements. The contact angle of water on EG films was measured under the ambient environment. The measurements were carried out on a two axis goniometer. Samples were horizontally mounted on the goniometer and a 5 μl drop was deposited on the sample's surface with an Eppendorf pipette. Determination of the contact angles was performed using the ImageJ software package fitting an ellipsoidal shape to the digital images of sessile drops. The uncertainty of the contact angle measurements varies sample by sample, but are no larger than 4°.

**b. Interfacial structure probed using high resolution X-ray reflectivity:**

The water-EG interface was probed by measuring the specular X-ray reflectivity (XR) signal at the 6-ID and 33-ID beamlines of the Advanced Photon Source. Application of XR techniques to liquid/solid systems and measurement procedures are described in detail in previous publications [22, 23]. The XR scattering geometry is illustrated in Fig. 2. Thin film sample cell and a Roper CCD X-ray detector were mounted on a six-circle goniometer (a Huber psi-C diffractometer at 6-ID and a Newport Kappa diffractometer at 33-ID). The incidence beam with a typical flux of around $10^{12}$ photons/s at the wavelength $\lambda$ = 0.9501 Å was reflected from the sample. The beam dimension was defined by a pair of slits [0.05-0.4mm (vertical) × 0.5-2mm (horizontal)]. Specular Bragg rod was recorded as a function of the vertical momentum transfer, $Q = (4\pi/\lambda)\sin(2\theta/2)$ (where λ is the X-ray wavelength, and 2θ is the scattering angle). This is also written in terms of the 6H-SiC reciprocal lattice index L in reciprocal lattice units (r.l.u.) as L = $Qc_{SiC}/2\pi$, where $c_{SiC}$= 15.12 Å is the vertical lattice spacing of the 6H-SiC substrate. The reflectivity signal was separated form background and diffuse x-ray scattering contributions in the CCD images to extract the absolute X-ray reflectivity for data analysis, following the procedure outlined previously [24].

**c. Analysis of X-ray reflectivity data:**

Typical XR data for EG films in contact with water are shown in Fig. 3(A) for a zero-layer graphene (ZLG) sample which consists of a $6\sqrt{}\times 6\sqrt{}R30°$ reconstructed carbon buffer layer ($G_0$) commensurate with SiC [25] and for bilayer graphene (BLG) with two partial carbon layers on top of $G_0$. In this work, the Fienup-type algorithm for one-dimensional X-ray imaging of interfacial structures at atomic scale [26] was applied to generate a satisfactory initial structural model as the starting model for subsequent model dependent non-linear least-squares fitting. The structural model consists of the ideal substrate with 6 SiC bilayers for each unit cell, the interfacial region including three top bilayers of the relaxed SiC, EG layers, ordered adsorbed water molecules, and bulk water expressed by a distorted layered-water model [27]. The multilayer morphology of the EG films (as shown in Fig. 1) indicates that the graphene layer that is locally in contact with water is found at different surface heights due to the incomplete graphene layer coverage. The partial occupation factors for each layer in the multilayer EG film structure is included in the analysis to represent the contribution of each interfacial water layer to the total structure factor. The total structural factor of the water/EG/SiC system can be expressed as

$$F_{tot} = F_{SiC}F_{CTR} + F_{INT} + \sum_{i=1}^{n}(occ_i - occ_{i+1}) \cdot F_{water,i} \tag{1}$$

which incorporates terms derived from the bulk SiC unit cell ($F_{SiC}$), the crystal truncation rod form factor ($F_{CTR}$), and the interfacial structure ($F_{INT}$) [22]. $F_{INT}$ includes the relaxed topmost SiC bilayers, all EG layers and adsorbed water layers. $F_{water,i}$ represents the layered interfacial water (extending to bulk water) contribution above each EG layer exposed to fluid water, whose layer occupancy is $occ_i$. A full electron density profile describing the system can be constructed directly from the structural model after convolution with the experimental resolution of the data. Important parameters of interest for the interfacial water structure above the different EG films obtained from model-dependent best-fits of specular XR data are shown in Table 1.

The goodness of fit to XR data are quantified by both a $\chi^2 = [\Sigma_k(I_k\text{-}I_{cal,k})^2/\sigma_k^2]/(N\text{-}N_p)$, which guides the fitting algorithm, and the R-factor ($R= \Sigma_k |(I_k\text{-}I_{cal,k})/I_k|/N$), where N and $N_p$ are the numbers of data points and parameters used in the model-dependent fitting, respectively, $I_k$ and $I_{cal,k}$ are the measured and calculated reflected intensities, respectively, and $\sigma_k$ is the uncertainty in the kth data point. Derived errors in the optimized fitting parameters are shown as 1σ uncertainties.

## B. Computational Methods

**a. Classical molecular dynamics simulations:**

Classical molecular dynamics (CMD) was used to determine the qualitative properties of pure water in contact with graphene sheets, represented here as mesoscopic slit pores. The simulations used the program Lammps [28] with a 1 fs time step. Long-range sums were computed using the Ewald method [29]. The box was periodic in all three dimensions, and contained a total of 1200 carbon atoms, 555 water molecules in a slit size of 4 nm. First, a 100 ps equilibration NVT (fixed number, volume, and temperature) was run, with the temperature gradually increased from 0 K to 300 K. Then a 200 ps NPT (fixed number, pressure, and temperature) simulation was performed at 0 GPa (P ~1 bar) and 300 K to ensure that the water would be at the appropriate density for the surface conditions. For all of the simulations, the box had the approximate x and y dimensions of 24.67Å and 21.37Å within the carbon plane and of 56.67Å perpendicular to it. While all three boundaries were allowed to vary (< 1Å), the rigidity of the carbon sheets prevented significant variation in the x and y directions. Finally, a 10 ns NVT simulation (T = 300 K) was performed of which 3 ns was used for supplemental equilibration. For the remainder, the position data was sampled and analyzed. The carbon atoms were frozen during this portion of the simulation [30]. Further details on the simulation methodology can be found in the reference 31 and references therein. Parameters for atoms O and H were taken from the ClayFF work and papers references therein [32-34]. The ClayFF force field uses the flexible, simple-point charge (SPC-Fw) water molecule. The carbon force field parameters were taken from the original Amber force field set [35]. All of these parameters are summarized in Table 2. A separate code was used to determine the atomic density profile perpendicular to the surface from the trajectory files [28]. The simulation box and the resultant number density profiles of respective atoms as a function of axis distance are illustrated in Fig. 4 (A) and (B).

In order to study the behavior of water in contact with graphene, we also performed isobaric-isothermal molecular dynamics simulations (another form of CMD) of pure water, described by the SPC/E model [36], surrounding a free-standing plate at ambient conditions, as demonstrated in Fig. 4 (C) and (D). The isobaric-isothermal molecular dynamics simulation of water was carried out according to our own implementation of a Nosé-Poincare symplectic integration algorithm [37, 38] with a time-step of 2.0 fs. The fluid environment consists of pure water and comprises a total of 2048 particles at T = 298 K and P = 1 atm within a tetragonal, $L_z = 2\ L_x = 2\ L_y$, simulation box subject to 3D periodic boundary conditions. A free standing graphene plate is immersed in the center of the box while kept fixed in space during the simulation. The graphene plate comprises 136 carbon sites explicitly described as Lennard-Jones spheres in the X-Y plane, characterized by $\varepsilon_{CC}/k$ = 28 K and $\sigma_{CC}$ = 3.40 Å [39, 40] and an adjacent carbon-carbon distance of 1.42 Å, i.e., 17.04Å by 18.44Å, where these dimensions are always smaller than the X-Y dimensions of the fluctuating system volume. The corresponding water-graphene Lennard-Jones unlike pair interactions are described by the Lorentz-Berthelot combining rules. Consequently, the environment in the two sides of a free standing graphene plate is able to exchange species with the surrounding, *i.e.*, behaving effectively as a grand canonical (open) system. In other words, this simulation box scheme allows us to analyze simultaneously the interfacial fluid behavior in equilibrium with its corresponding bulk at precisely the same global state conditions. For additional details on the simulation approach and results the reader should refer to reference [41].

**b. *Ab-initio* molecular dynamics simulations:**

*Ab-initio* density functional molecular dynamics simulations, which adequately capture the interfacial chemistry, were performed for water on epitaxial graphene with a few different surface modifications to test all relevant hypotheses. The simulations were done using a Perdew-Burke-Ernzerhof generalize-gradiant approximation (PBE-GGA) functional and a projector-augmented-wave (PAW) potential as implemented in VASP [42-44]. All figures shown in the main text and this document correspond to simulations at T = 600K, fixed using a Nose thermostat, and a water density of ~1g/cc. A time-step of 0.5 fs was chosen. The Water/G system had one hundred water molecules with the initial configuration coming from TIP3 water. The system was simulated with and without van der Waals potential at T = 400K for 43 and 32 ps respectively. Effect of dispersive forces between water and graphene was explored by adding a Grimme type potential to the total Hamiltonian [45]. No change in the water-graphene first-neighbor peak position/height was seen, although beyond the first-neighbor, where water is essentially close to the bulk density, the structure is affected by inclusion of van der Waals forces. Because of the negligible effect of dispersive forces on the water-graphene interaction, we use the pure PBE-DFT for all other calculations. A T = 600K simulation was subsequently carried out for another 11.5 ps.

To model the SiC-6H surface, a 6-bilayer SiC was constructed (essentially the primitive cell of SiC-6H) with one end terminated with Si and the other end with C. A 2×2×1 supercell of this structure was then capped with a single graphene layer (which is 2√3×2√3 R30º) on each end, forming the well-known buffer layer graphene. The buffer layer on the Si-terminated side (henceforth referred to as $G_0$-layer) is the main focus of this study. The relaxed surface had a ~8% strain in the $G_0$ surface layer with a corrugation of ± 0.3-0.4 Å. To simulate the real buffer-layer corrugation, a relaxation study of the fully commensurate 6√3×6√3 R30º structure of the SiC-6H and graphene interface was performed. The resulting corrugation was similar to that of the 2√3×2√3 R30º graphene. The $G_0$ layer had $sp^2$ and $sp^3$ carbons with the $sp^3$ carbons bonded to the underlying Si atoms with a bond-length of about ~2.02 Å. The structure is in good agreement with published DFT results [46, 47]. A box size was chosen to accommodate 72 water molecules between the two carbon-terminated interfaces, with a density of 1g/cc, and with an initial hexagonal-ice structure and was melted at T = 300K for 5ps. Table 3 shows the run-lengths for the different cases. All data analysis was performed after an initial transient of ~2-3ps.

## IV RESULTS AND DISCUSSIONS

### a. Observations of anomalous interfacial water structure on epitaxial graphene:

The derived best-fit electron density profile of the BLG sample in contact with water is shown in Fig. 3(B). $G_0$ is strongly bound to SiC with a nearly complete coverage at a height $d_0$ = 2.08 ± 0.01 Å above the uppermost Si layer, which is smaller than that between subsequent layers (very close to the bulk graphite value of 3.35 Å for $G_0$-$G_1$ and $G_1$-$G_2$ for BLG). $G_0$ also has a larger intrinsic root-mean-square width than that of $G_n$ (*i.e.*, for n>0) indicating a significant corrugation. Both findings are consistent with most recent DFT calculations, surface X-ray diffraction and scanning tunneling microscopy studies [48, 49]. The decreasing coverage of subsequent EG layers implies that interfacial water structures are formed separately in contact with $G_0$, $G_1$, $G_2$, etc. In contrast, the ZLG has only a $G_0$ layer with a height similar to that for $G_0$ in the BLG sample.

The complex water structures above each of the exposed EG layers are shown as separate profiles [Fig. 5(A)]. The peak at the origin represents the outermost graphene layer that interacts with water, and each of the vertical density profiles are normalized by the partial coverage of the particular EG layer which is exposed to water. Distinct peaks of adsorbed water are observed above respective EG layers, suggesting significant water layering near graphene [3, 5, 7]. Using the same structural model to analyze the data, we find for BLG that the first interfacial water layer is at an average height of 2.28 ± 0.02 Å above $G_0$, significantly smaller than that observed above $G_n$ (3.19 ± 0.02 Å). The relatively low fraction of $G_0$ (18 %) that is in direct contact with water in BLG explains why its macroscopic contact angle is close to bulk graphite (93° ± 3°). Separate measurements reveal an average water-$G_0$ distance of 2.44 ± 0.03 Å for the ZLG sample (with almost full exposure of $G_0$ to water) [Fig. 5(B)], concomitant with the significantly reduced water contact angle (73 ± 4°) and confirming the anomalous water structure that is found in contact with the $G_0$ layer. This reduced graphene-water separation further suggests that the strained and corrugated buffer layer ($G_0$) interacts, on average, more strongly with water as compared with intrinsic graphene ($G_n$).

Measurements on EG films with varying thicknesses reveals a range of contact angles between that of the ZLG and BLG. Parallel XR results show that the change in contact angle is controlled by the structure of the EG film, specifically the fraction of the EG layer that is terminated by $G_n$ [here expressed as $OCC_{Gn\text{-}W}$; see Fig. 5(B), inset]. The water-exposure-coverage was obtained independently from each EG sample by XR. Once the water-exposure-coverage of $G_n$ approaches unity, the contact angle becomes independent of numbers of graphene layers (93 ± 3°). Given the different interfacial water structures, the relative hydrophobicity of these EG films can therefore be envisioned as being mediated by the relative portion of $G_n$ vs. $G_0$ patches in direct contact with water.

**b. Insights from molecular dynamics simulations:**

The reduced hydrophobicity of $G_0$ with respect to $G_n$ can be due to multiple mechanisms, such as the inherent corrugation of $G_0$ [51], the underlying SiC substrate [52], and Stone-Wales [53] or vacancy [8] defects. Defects may introduce sp bonded carbons on the surface and open larger rings (slit-pores) compared to the otherwise unique hexagon ring on defect-free graphene, acting as potential sites for splitting water [54].

To test these different possibilities, we performed simulations for a variety of mechanisms using different levels of theory. Firstly, CMD simulations with empirical water force fields were performed by adopting two distinct configurations: a mesoscopic graphite slit pore filled with water and a freestanding graphene surrounded by water (see Fig. 4). The primary features in the density profiles from both CMD simulations [Fig. 5(A)] of the interfacial water above an ideal graphene sheet are two distinct layers of water peaked at 3.2 Å and 6.1 Å, which are consistent with previous CMD studies [3, 7] and in agreement with our experimentally determined water structure above $G_n$. This confirms that water interacts weakly with intrinsic graphene and bulk graphite through van der Waals (vdW) interactions [6].

$G_0$ has strong covalent interactions with SiC and such chemical differences may not be reflected in the CMD force fields [55]. *Ab initio* MD (AIMD) simulations within the DFT framework were performed for water on epitaxial graphene with a few different surface modifications to test all relevant hypotheses. The oxygen number-density profiles for different simulations at T = 600K are illustrated in Figure 5(D) (each normalized to the bulk density of water for the particular simulation). A higher temperature was used to allow faster equilibration of water within simulated time scales. Lower temperature studies at T = 400K for long times (~35ps) were also performed for water on graphene and agree with the T = 600K results. These AIMD results show that interfacial water behaves similarly on a free-standing graphene (FSG), whether it is flat (water/FSG), or corrugated with amplitudes of ~ 0.3-0.4 Å (water/$FSG_0$) or decorated with a single vacancy defect (water/FSG-1vac). Notable features are a first water-oxygen peak position at 3.3 Å and a rapidly decaying density-oscillation to the bulk water density above ~ 7.0 Å from the interface. The density profiles from our FSG simulations are consistent with a previous AIMD work [5] as well as with our CMD results, indicating that water-FSG interaction is indeed hydrophobic and only weakly sensitive to the structure of the FSG layer.

In contrast to FSG, the epitaxy of graphene on SiC ($G_0$/SiC) down-shifts the first water-oxygen peak (on average by ~0.2 Å) and increases by 23% the number of water molecules within 3.3 Å of the surface, indicating a stronger interaction of water on $G_0$/SiC. This change is due to the presence of $sp^3$ and $sp^2$ carbons on $G_0$/SiC which are absent on FSG or $FSG_0$. To model the different bonding environments resulting from surface defects, we introduce a single vacancy defect by removing a $sp^2$ carbon on $G_0$/SiC and allowing the surface to reconstruct. The reconstructed surface has sp, $sp^2$ and $sp^3$ bonded carbon atoms with a ~ 97% carbon occupancy, comparable to experiment. AIMD simulations show that this vacancy defect (W/$G_0$-1vac/SiC) allows water molecules to move closer to $G_0$. In addition, evidence for chemical bonding is seen between the closest lying water molecule and $G_0$ resulting in a significant charge-transfer of about 0.03 $e^-/Å^3$ [Fig. 6(A)]. This effect is due to the presence of the SiC surface. This localization of water is dynamic, with multiple water monomers moving in and out of the pore during the simulation, and is localized to the vacancy. The number of water molecules within 3.3 Å from the interface is the same with/without a vacancy on $G_0$/SiC, which accounts only for a fraction of the first water layer shift revealed from XR analysis (0.75 Å). This quantitative disagreement implies that there is an additional surface feature in the real system absent in the vacancy-only AIMD model.

Defect sites are more reactive for OH adsorption than pure graphene [52, 54]. We find that the adsorption energy of a hydroxyl group on a $sp^2$ site of an ideal $G_0$/SiC surface is ~2.0 eV/OH, while that on a free-standing graphene is ~1.3 eV/OH. We simulate the effect of water splitting (W/$G_0$-OH+H/SiC) by replacing a water molecule with co-adsorbed OH and H on two $sp^2$ carbons of $G_0$/SiC. The z-profile of oxygen from AIMD of water shows a strong peak at 3.4 Å with a shoulder at 3.1 Å. In addition, there is an extra peak at ~ 1.8 Å due to the covalently bound hydroxyl on the surface.

Hence, the water anomaly on $G_0$ observed by XR can be reasonably quantified when incorporating the effect of both defects and surface-hydroxyls. The combination of these two features, as illustrated in Fig. 5(C), is seen to agree qualitatively with the experimentally determined profile when we allow the first water layer to described by a two-peak model for the interfacial water structure rather than a single broadly distributed Gaussian as shown in Fig. 5(B). That is, the shift in the average height of the first interfacial water layer is shown to be due to a change in the interfacial water structure suggested by the DFT theory results (this more complex optimized structural model does not alter our central observation that the average height of the interfacial layer at $G_0$ is substantially smaller than that observed at $G_n$). The observed changes of interfacial water near $G_0$ is also manifested as the perturbation of the average H-O-H bond-angle in water molecules for heights of 1.5-4 Å from the interface [Fig. 5(D), inset], which is seen as large bifurcations from the otherwise monotonic decrease observed on FSG, consistent with having two types of carbon ($sp^2$ and $sp^3$ carbons) on the surface. Vacancy defects alter this bifurcation and make the perturbations extend to larger heights.

To better understand the site-specificity of water-graphene interaction, we investigate the lateral (2D) surface-density profiles of water-oxygen within a distance of 2 Å < z < 3 Å from the surface for the four selected cases (Fig. 7). The lateral water distribution is laterally uniform on $FSG_0$, but it becomes strongly non-uniform with the insertion of SiC, with the water molecules diffusing over long-distances [Fig. 7(B)] to $sp^3$ sites. Polar water molecules are flexible enough to reorient according to the underlying corrugation in the surface electrostatic potential energy, preferring sites with a lower potential energy expected on the covalently bonded $sp^3$ sites. This is clearly seen from the charge-density isosurfaces [Fig. 6(B)] which show enhanced charge density near $sp^2$ sites compared to $sp^3$ sites.

Further, introduction of a vacancy on $G_0$ [Fig. 7(C)] draws water deeper into the pore and, hence, closer to the solid interface. In fact, most of the water immediately adjacent to the graphene surface prefers to be close to the defect site. The addition of a single hydroxyl group [Fig. 7(D)] appears to push water away from the hydroxyl-site in this layer while attracting water in the next layer. All this leads to a strong variability of the first-monolayer of water with the water/carbon distance reduced because of water seeping into the pores [Fig. 7(C)]. This not only explains the observed interfacial water behavior, but also suggests the possibility of new routes to control flow of water in nano-fluidic applications.

### c. Implications for the hydrophobic gap: Determination of the intrinsic width

A critical open question concerns our finding that a nominal depletion region is revealed between intrinsic graphene ($G_n$) and the first interfacial water layer, as shown in Fig. 5(A). This depletion region is predicted by CMD and AIMD simulations and in agreement with a previous study [5]. Experimentally, this density gap is inevitable when observations are made with the sub-Ångstrom (< 0.5 Å) resolution of XR measurements. The observed profile suggests a nominal gap of ~1 Å width, after the radii of carbon (0.77 Å) and water molecule (1.44 Å) are included. The intrinsic gap width can also be determined directly from the density profile, as shown in Fig. 8(A). The density profile for the graphene layer is obtained from XR analysis. The density profile for both oxygen and hydrogen atoms in the aqueous phase are extracted from CMD simulation results. Therefore, the gap width starts from the tail of the density peak for carbon to the closest H-density profile, where no carbon valence electrons or H-protons contribute to the depletion gap. As indicated in the figure, the intrinsic gap width obtained is very close to 1 Å, which is consistent with the value obtained by simply accounting for the steric constraints of each component.

However, a more conservative gap width would be obtained if we take into account also the van der Waals (vdW) interactions of the graphene layer. Fig. 8(B) illustrates a generic way to determine the intrinsic gap width by counting the difference of the integrated density within the interfacial region between the case of an ideal water/solid interface (without gap) and the experimentally-determined or simulated density profile for water on intrinsic graphene. The interfacial region is bounded by a window defined by the vdW radius of a carbon atom and the density profile for bulk water above the interface. A positive definite integrated density difference indicates the existence of an intrinsic gap between two phases. This generic estimation is equivalent with the product of the depletion region width and the density deficit (δD·δρ), a more robust parameter to describe the net depletion of the density at a hydrophobic interface since the details of a hydrophobic gap are often not intrinsically resolved by most low angle reflectivity measurements [56-58]. From Fig. 8(B), the obtained integrated density difference for the case of water on intrinsic BLG sample is $\delta D \cdot \delta\rho = 0.0623$ $e^-/Å^2$. If the density deficit of water is assumed to be 0.33 $e^-/Å^3$ [fully depleted, consistent with the zero electron density region that is observed in Fig. 8(A)], the intrinsic gap width is determined to be 0.2 Å. The gap is substantially smaller than that inferred by previous studies of more highly hydrophobic interfaces, but is consistent with the idea of a gap whose magnitude is a function of the contact angle [56]. It is compelling to note that no significant depletion is obtained by the same method (using the density profile from AIMD results) for the case of water on the epitaxial buffer layer ($G_0$) (with a derived value of ~ -0.02 $e^-/Å^2$), suggesting that no such gap exists between water and intrinsically defective $G_0$ due to the increased chemical bonding of water with the defective sites. This reduction in the density gap with respect to intrinsic graphene is in a good agreement with the observation of the decrease of the contact-angle with increased water/$G_0$ coverage. These results therefore provide more insights into the structure of water at hydrophobic interfaces that were not intrinsically resolved in previous resolution-limited studies [56-58] and avoid the uncertainties due to the invisibility of hydrogen atoms to X-rays [59].

## V. SUMMARY

In summary, we have explained the controllable interaction of water with epitaxial graphene films of different thickness. An intrinsic depletion region between water and $G_n$ was found to be greatly reduced above $G_0$. The macroscopic wettability of multilayer graphene is controlled by the local structure and coverage of micron-scale hydrophilic and hydrophobic patches, consisting of intrinsic graphene ($G_n$) and the defective $G_0$/SiC, respectively. Simulations show that this control is achieved by three factors: the inclusion of $sp^2$ vs. $sp^3$ character of the $G_0$ layer due to its epitaxy with SiC; the inclusion of covalent interactions between water and defective epitaxial-graphene; and the inclusion of functionalized surface groups (e.g., adsorbed hydroxyls) leading to lateral heterogeneity of interfacial water. Our integrated modeling and experimental study provides unique insights into the chemical differences between intrinsic graphene ($G_n$) and the epitaxial buffer layer ($G_0$), which is fundamental to understanding and controlling the interactions of aqueous and non-aqueous fluids in electrochemical energy storage systems.

## ACKNOWLEDGEMENTS

We thank Sang Soo Lee for his advice and assistance in X-ray experiments and data analysis. This material is based upon work supported as part of the Fluid Interface Reactions, Structures and Transport (FIRST) Center, an Energy Frontier Research Center funded by the U.S. Department of Energy (DOE), Office of Science (SC), Office of Basic Energy Sciences (BES) under Award Number ERKCC61. Use of the beamlines ID6 and ID33 at the Advanced Photon Source was supported by DOE-SC-BES under contract DE-AC02-06CH11357 to UChicago Argonne, LLC as operator of Argonne National

Laboratory. This research used resources of the National Energy Research Scientific Computing Center, which is supported by DOE-SC under Contract No. DE-AC02-05CH11231. V.P. acknowledges financial support by the Alexander von Humboldt Foundation. Raman spectroscopy was carried out on equipment of the W. M. Keck Institute for Attofluidic Nanotube-Based Probes at Drexel University.

* hzhou@anl.gov

†ganeshp@ornl.gov

‡fenter@anl.gov

## Figure and Table Captions:

Table 1: Parameters of interfacial water structures above epitaxial graphene obtained from model-dependent best-fit of specular X-ray reflectivity data.

Table 2: Force field parameters for classic molecular dynamics simulations.

Table 3: Details of *ab-initio* molecular dynamics simulations for the different geometries analyzed at T = 600K.

Figure 1: Growth and characterization of EG on SiC. (A) EG is synthesized by vacuum graphitization of a SiC wafer. A zero-layer graphene (ZLG) and few-layer graphene (FLG) can be obtained from Si-face growth. (B-D) Raman spectra serves as a fingerprint to characterize the graphene thickness; (B) A wide range Ramam spectra identifying both SiC and epitaxial graphene excitation bands; (C) Detail of the region between 1300 and 1800 $cm^{-1}$; (D) Detail of the region between 2500 and 3200 $cm^{-1}$. AFM image showing the (E) height and (F) derivative contrast image. In both figures, the solid circle represents the first monolayer graphene and solid diamond for the second monolayer graphene with an incomplete coverage. The scale bar is 1 μm.

Figure 2: Experimental setup and scattering geometry for high resolution X-ray reflectivity of epitaxial graphene on SiC in contact with bulk water. $K_i$: X-ray incidence vector; $K_f$: X-ray scattering vector; Q: scattering momentum transfer $K_f - K_i$.

Figure 3: (A) XR data from ZLG (diamonds) and BLG (circles) film in contact with liquid water and best fits, as compared with the calculated XR for an ideal SiC substrate.

(B) The density profile generated from the structural model for BLG. Red solid line: relaxed SiC and EG layers; Red dashed line: adsorbed water above respective EG layers; Yellow (Cyan, Magenta) dashed line: DLW above $G_0$ ($G_1$, $G_2$); Black solid line: total electron density of water. Purple shaded band: ±1σ vertical uncertainties for water density. $d_1 = 3.43 \pm 0.02$ Å; $d_2 = 3.35 \pm 0.09$ Å. Inset: Schematic drawing of the structural model. Blue: Si; Brown: C in SiC; Red: C in graphene; Black: O; White: H.

Figure 4: Classical Molecular dynamic (MD) simulations of graphene in contact with water. (A) Classic MD simulation of water under confinement between graphene sheets (finite-size slit-pore). The water structure resembles with the case of bulk water for the slit size of 4 nm in this study. (B) The resultant number density profiles of respective atoms as a function of axis distance. (C) Isobaric-isothermal MD simulation of water at the fluid-graphene interfaces (a free-standing sheet). (D) The number density profiles of respective atoms as a function of axis distance.

Figure 5: Interfacial water structures for (A) BLG and for (B) ZLG, as compared to classic MD simulations. AdsW--$G_0$($G_n$): the adsorbed water (AdsW) above $G_0$ ($G_n$); DLW: the distorted layered-water for bulk. Left inset in (A) and (B): the contact angle for BLG and ZLG. Right inset in (B): the dependence of the water-exposure-coverage for $G_n$ patches on contact angle and the simply calculated one by Cassie's Law [50] (dashed line). (C) The density profile from the refined structural model for ZLG, as guided by AIMD simulations. (D) Oxygen number-density profiles for AIMD simulations of water on graphene under different interfacial configurations. Inset: average H-O-H bond-angle profiles. Shaded area: the integration volume for the 2D-plots in Fig. 3.

Figure 6: (A) Isosurfaces for charge-density differences with and without the $W_1$ water molecule for an AIMD snapshot of water/$G_0$(1vac)/SiC run. Red is positive and blue is negative isosurface at a value of 0.03 $e^-/Å^3$. Increased charge-transfer is seen between water and $G_0$ near the vacancy site. (B) Charge density isosurface (magenta) with a value of 1.55 $e^-/Å^3$ from an AIMD snapshot showing the increased charge-density near $sp^2$ carbons compared to the $sp^3$ carbons on $G_0$/SiC.

Figure 7: Oxygen number density (in g/cc × 100) between 2-3 Å from the interface for (A) Water/$FSG_0$ (B) Water/$G_0$/SiC (C) Water/$G_0$(1vac)/SiC and (D) Water/$G_0$(OH+H)/SiC with the underlying $G_0$ shown as an inverted-grey-scale plot in the z-range [-0.5,0.5] Å about the average $G_0$ position. Our simulated supercell, is outlined in orange. $sp^3$ bonded carbon atoms appear in light grey color while $sp^2$ carbons have a darker shade, corresponding to their increased z-height. OH and H adsorption sites are schematically shown in (D). The long streaks are long diffusion paths of water over time.

Figure 8: Determining the intrinsic width of the hydrophobic gap for an arbitrary profile. (A) The nominal gap width determined from the zero-density region measured and simulated density profiles, including C, O and H atoms. (B) A generic way to determine the gap width. The interfacial region is bounded by the window, marked with dashed lines. $r_W$ is the radius of water molecule; $r_G$ is the vdW radius of a carbon atom.

**Table 1**

| Parameters \ Sample | EG4 (ZLG) | EG2 | EG22 (BLG) |
|---|---|---|---|
| No. of graphene layer | "0" | "0 + 2" | "0 + 2" |
| Occupancy for graphene layer | $G_0$: 1.01 ± 0.03<br>–<br>– | $G_0$: 0.95 ± 0.01<br>$G_1$: 0.66 ± 0.05<br>$G_2$: 0.15 ± 0.03 | $G_0$: 0.92 ± 0.01<br>$G_1$: 0.82 ± 0.01<br>$G_2$: 0.11 ± 0.04 |
| Intrinsic r.m.s. width for graphene layer (Å) | $G_0$: 0.24 ± 0.01<br>–<br>– | $G_0$: 0.20 ± 0.02<br>$G_1$: 0.10 ± 0.03<br>$G_2$: 0.10 ± 0.01 | $G_0$: 0.24 ± 0.01<br>$G_1$: 0.10 ± 0.01<br>$G_2$: 0.10 ± 0.01 |
| Last Si-C bilayer disp. from bulk (Å) | Si: 0.03 ± 0.01<br>C: -0.01 ± 0.02 | Si: -0.06 ± 0.01<br>C: 0.26 ± 0.03 | Si: 0.07 ± 0.01<br>C: 0.19 ± 0.04 |
| Last Si-C bilayer occupancy | Si: 1.02 ± 0.03<br>C: 1.06 ± 0.04 | Si: 0.97 ± 0.02<br>C: 0.89 ± 0.03 | Si: 0.71 ± 0.04<br>C: 1.24 ± 0.07 |
| $G_0$ to last Si dist. (Å) | 2.19 ± 0.01 | 2.37 ± 0.01 | 2.08 ± 0.01 |
| $G_1$ to $G_0$ (Å) | – | 3.49 ± 0.04 | 3.43 ± 0.02 |
| $G_2$ to $G_1$ (Å) | – | 3.20 ± 0.10 | 3.35 ± 0.09 |
| Adsorbed water to $G_0$ dist. $d_{gw}$ (Å) | 2.44 ± 0.03 | 2.26 ± 0.09 | 2.28 ± 0.02 |
| Adsorbed water to $G_n$ (n>0) dist. $d_{gw}$ (Å) | – | 3.07 ± 0.09 | 3.19 ± 0.02 |
| Weighted adsorbed water to graphene layer dist. $d_{gw}$ (Å) | 2.44 ± 0.03 | 2.79 ± 0.09 | 3.06 ± 0.02 |
| Adsorbed water to distorted layered water dist. $d_{OO}$ (Å) | $G_0$: 3.72 ± 0.07<br>– | $G_0$: 2.71 ± 0.15<br>$G_n$: 1.90 ± 0.15 | $G_0$: 3.97 ± 0.08<br>$G_n$: 3.14 ± 0.09 |
| Distorted layered water dist. $d_{LW}$ (Å) | 3.93 ± 0.45 | 3.18 ± 0.22 | 2.90 ± 0.18 |
| Distorted layered water width $\delta_o$ (Å) | 0.78 ± 0.08 | 0.64 ± 0.10 | 0.57 ± 0.04 |
| Total water thickness (μm) | 26 ± 2 | 38 ± 3 | 60 ± 2 |
| Robinson roughness parameters β | 0.02 ± 0.02 | 0 | 0.19 ± 0.01 |
| Fit goodness $\chi^2$ (R factor) | 5.5 (5.6 %) | 1.9 (4.8 %) | 1.5 (3.4 %) |
| Contact angle (°) | 73 ± 4 | 80 ± 4 | 93 ± 3 |
| Structure Model | Parameters of the adsorbed water above $G_0$ layer optimized independently; Identical parameters used for adsorbed water above subsequent $G_n$ layer (n>0) | | |

**Table 2.**

| Force Field Parameters | | |
|---|---|---|
| **Element** | **Mass/AMU** | **Charge/$e^-$** |
| Graphite C | 12.011 | 0.00 |
| Water O | 15.9994 | 0.82 |
| Water H | 1.00797 | 0.41 |
| **Lennard Jones 12-6 Parameters** | **Pair Coefficients** | |
| | **ε/kcal mol$^{-1}$** | **σ/Å** |
| Graphite C | 0.1200 | 3.2963 |
| Water O | 0.1554 | 3.1655 |
| Water H | 0.0000 | 0.0000 |
| **Bond Coefficients** | | |
| | **k/kcal Å$^{-1}$** | **r/Å** |
| Graphite C-C | 469.0 | 1.4 |
| Water O-H | 554.135 | 1.0 |
| **Angle Coefficients** | | |
| | **k/kcal degree$^{-1}$** | **θ/degrees** |
| Graphite C-C-C | 85.0 | 120.0 |
| H-O-H | 45.7696 | 109.47 |
| **Dihedral Coefficients** | | |
| | **k/kcal degree$^{-1}$ Phase Angular Freq.** | |
| Graphite C-C-C-C | 5.3 -1 2 | |

**Table 3**

| System | Number of atoms | Run length at T=600K in ps |
|---|---|---|
| Water/FSG | 360 | 11.5 |
| Water/FSG$_0$ | 280 | 13.3 |
| Water/G$_0$/SiC | 424 | 13.5 |
| Water/G$_1$/G$_0$/SiC | 488 | 6.2 |
| Water/FSG(1vac) | 359 | 8.8 |
| Water/G$_0$(1vac)/SiC | 423 | 9.9 |
| Water/G$_0$(OH+H)/SiC | 424 | 10.7 |

**Figure 1**

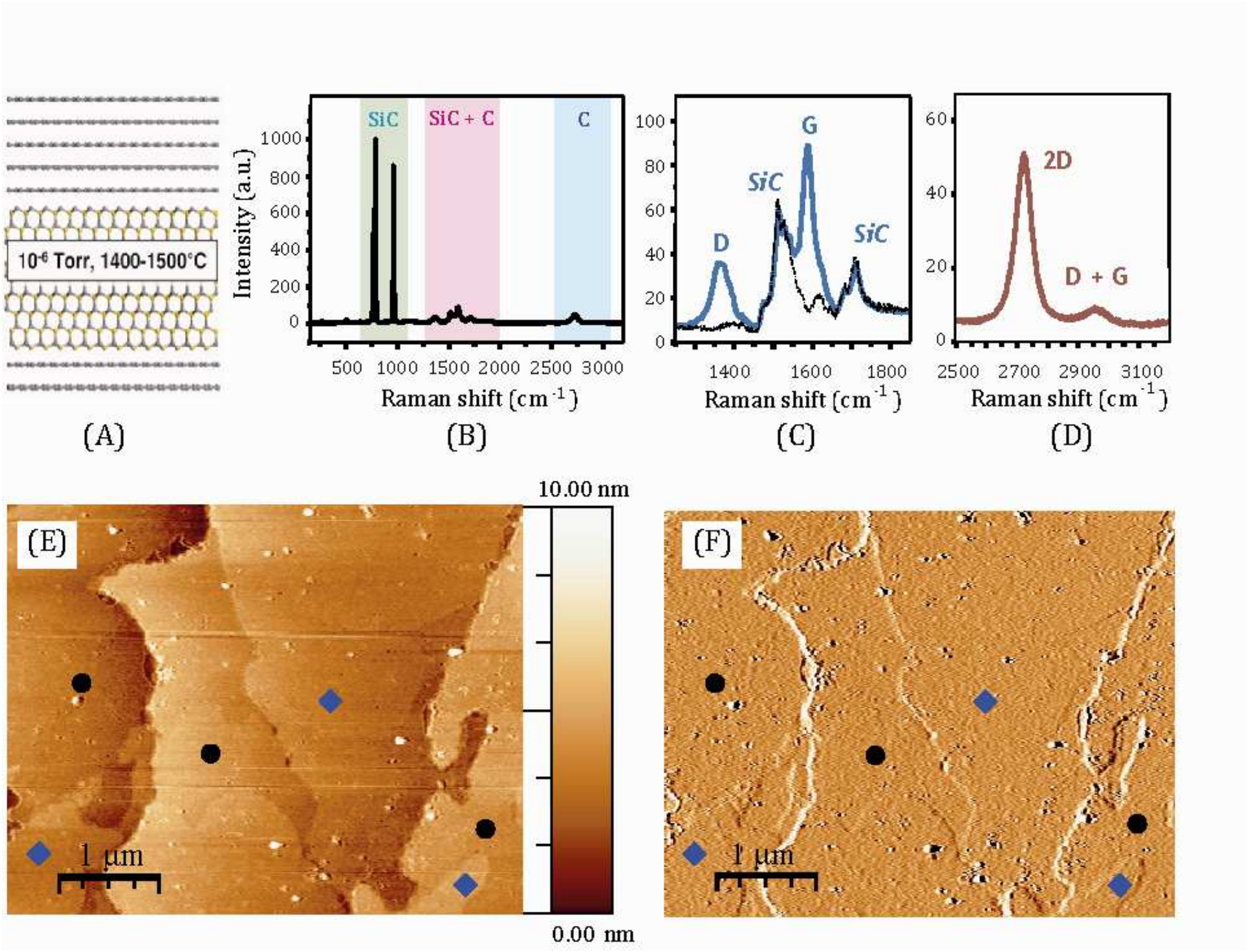

**Figure 2**

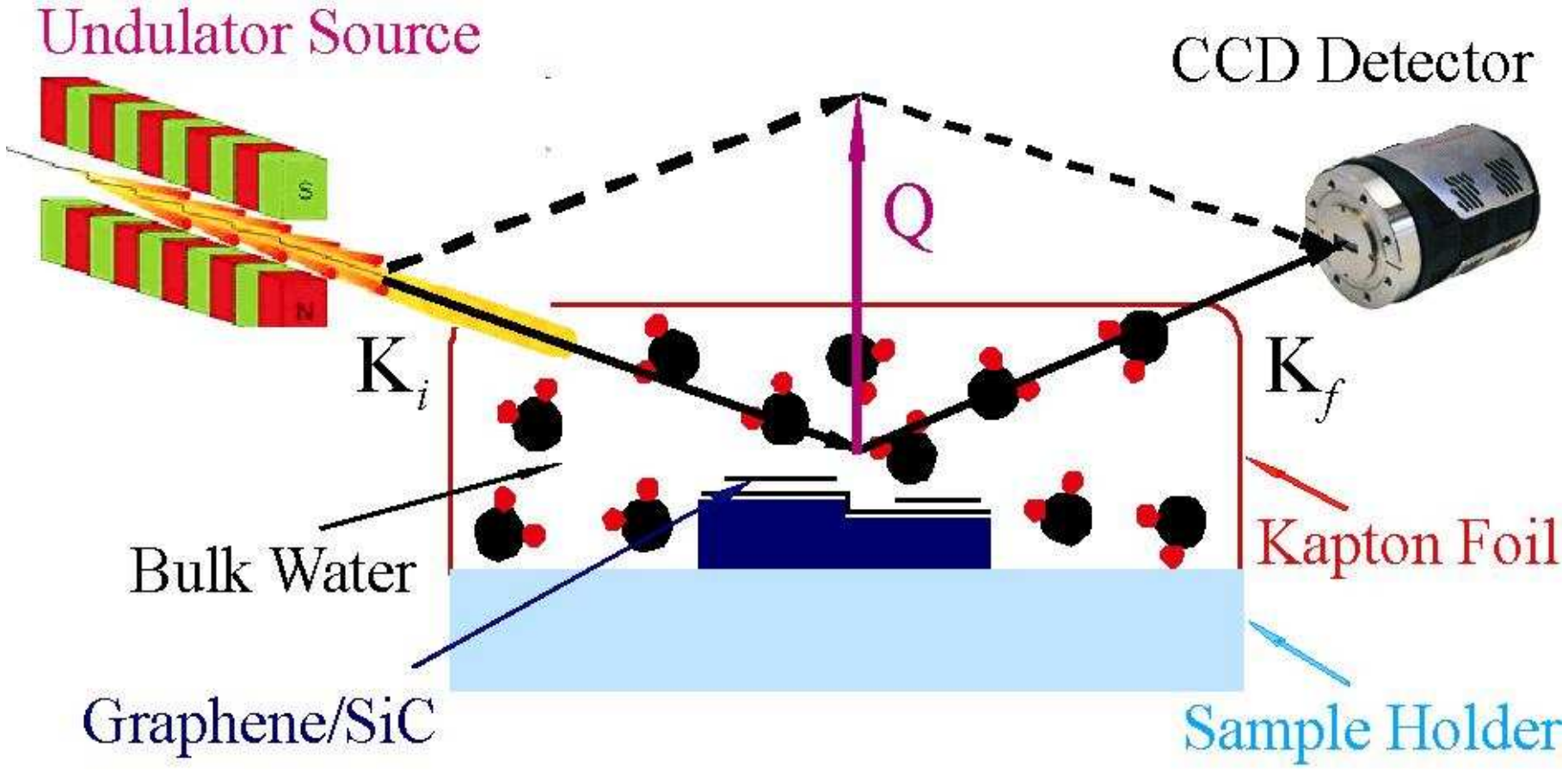

**Figure 3**

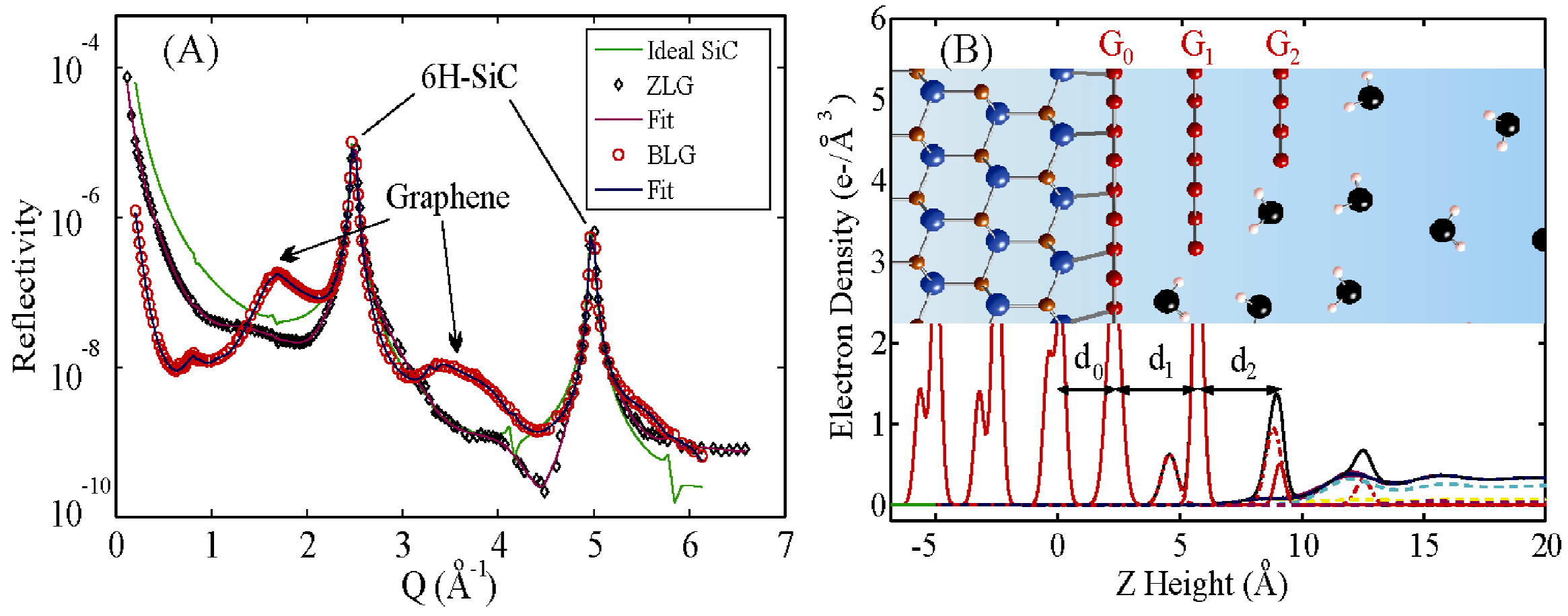

**Figure 4**

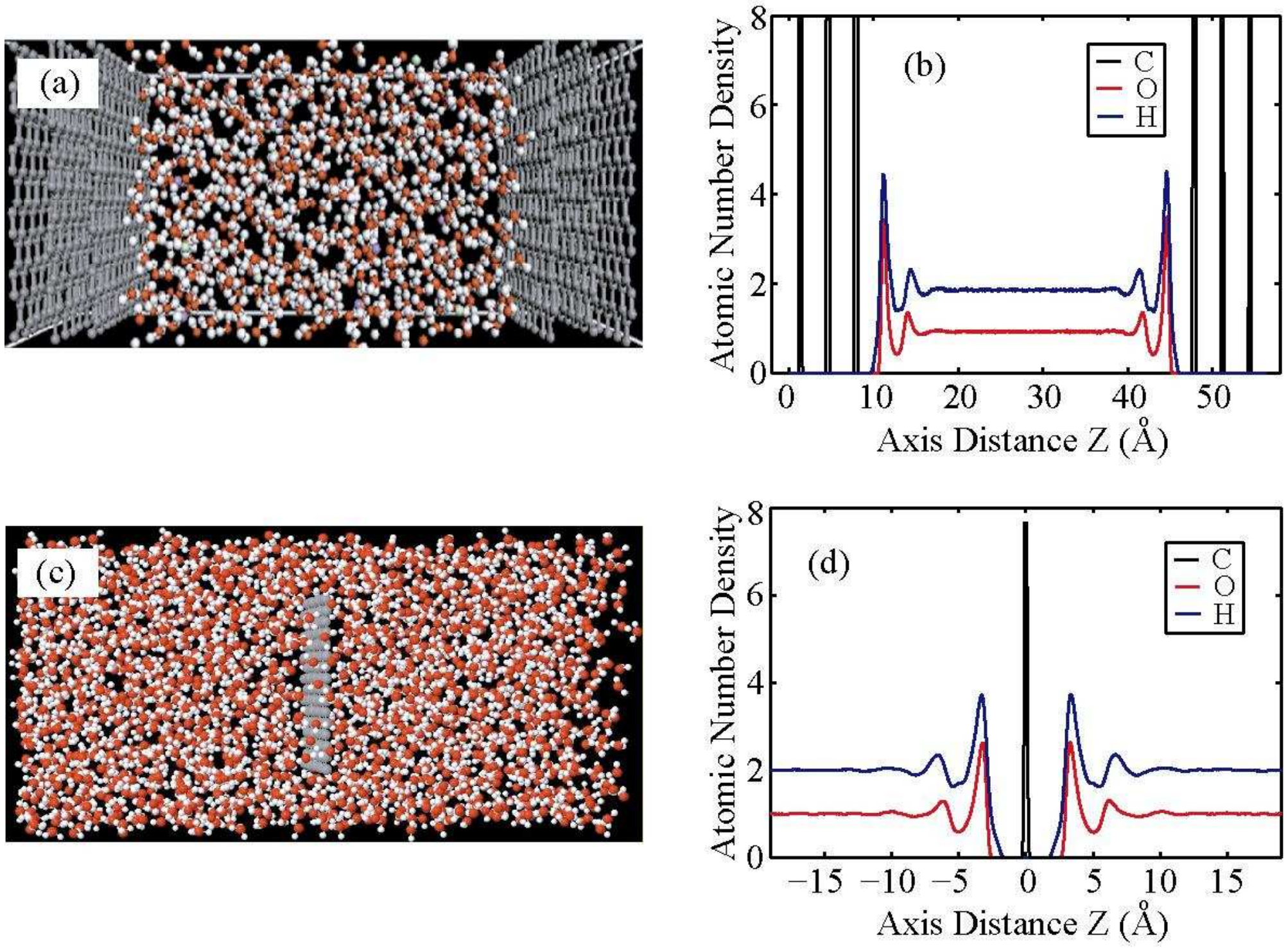

**Figure 5**

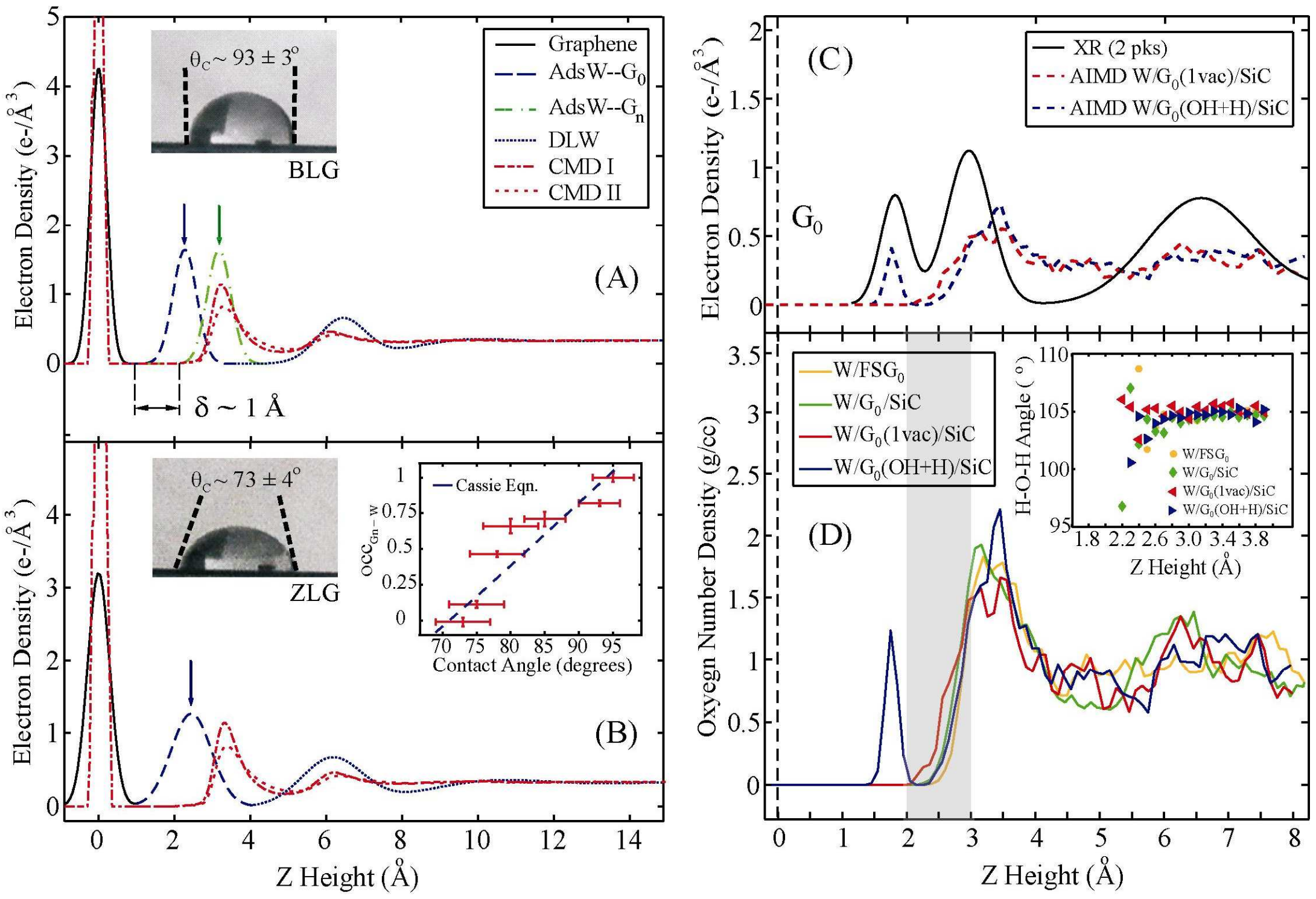

**Figure 6**

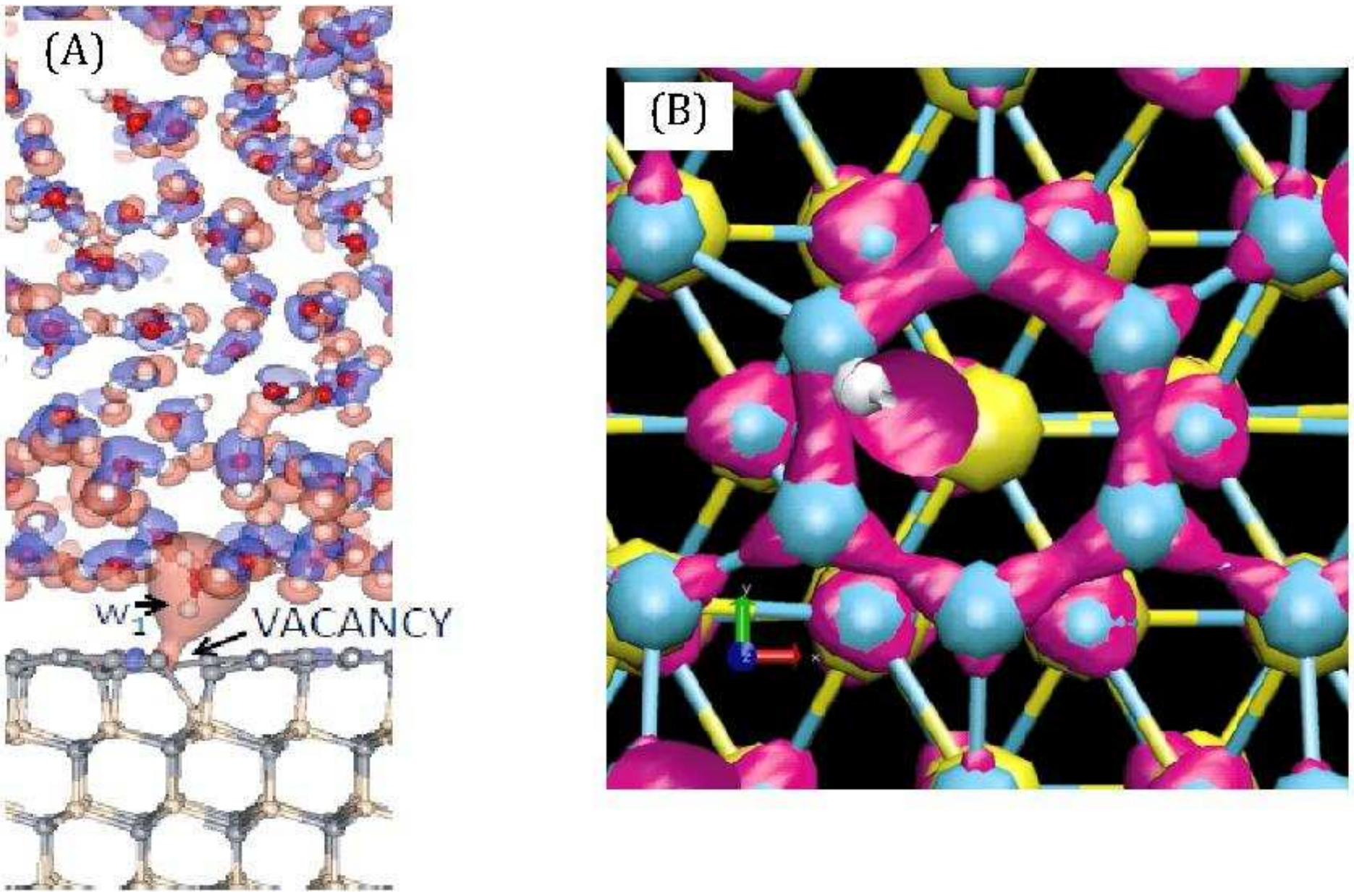

**Figure 7**

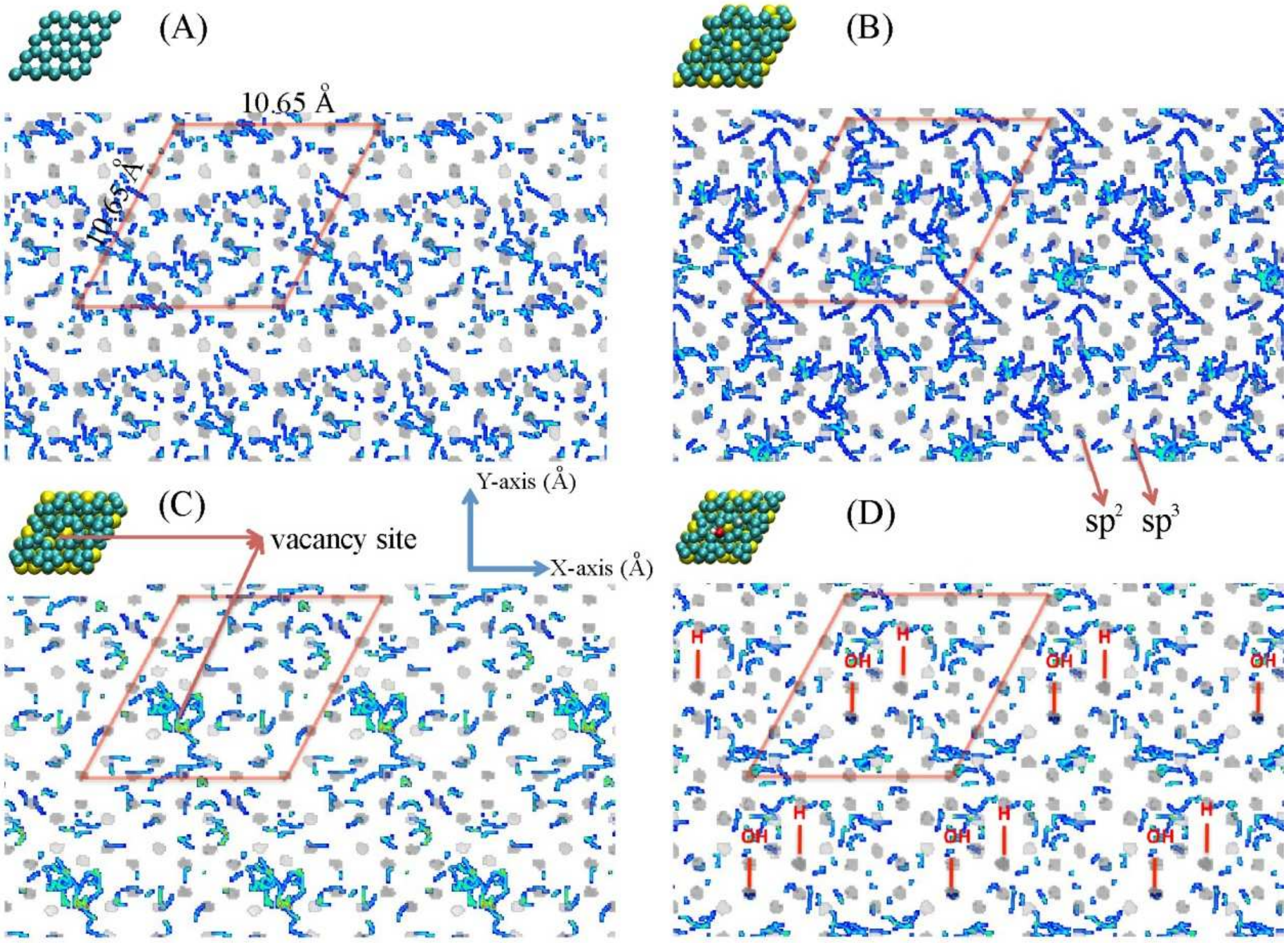

**Figure 8**

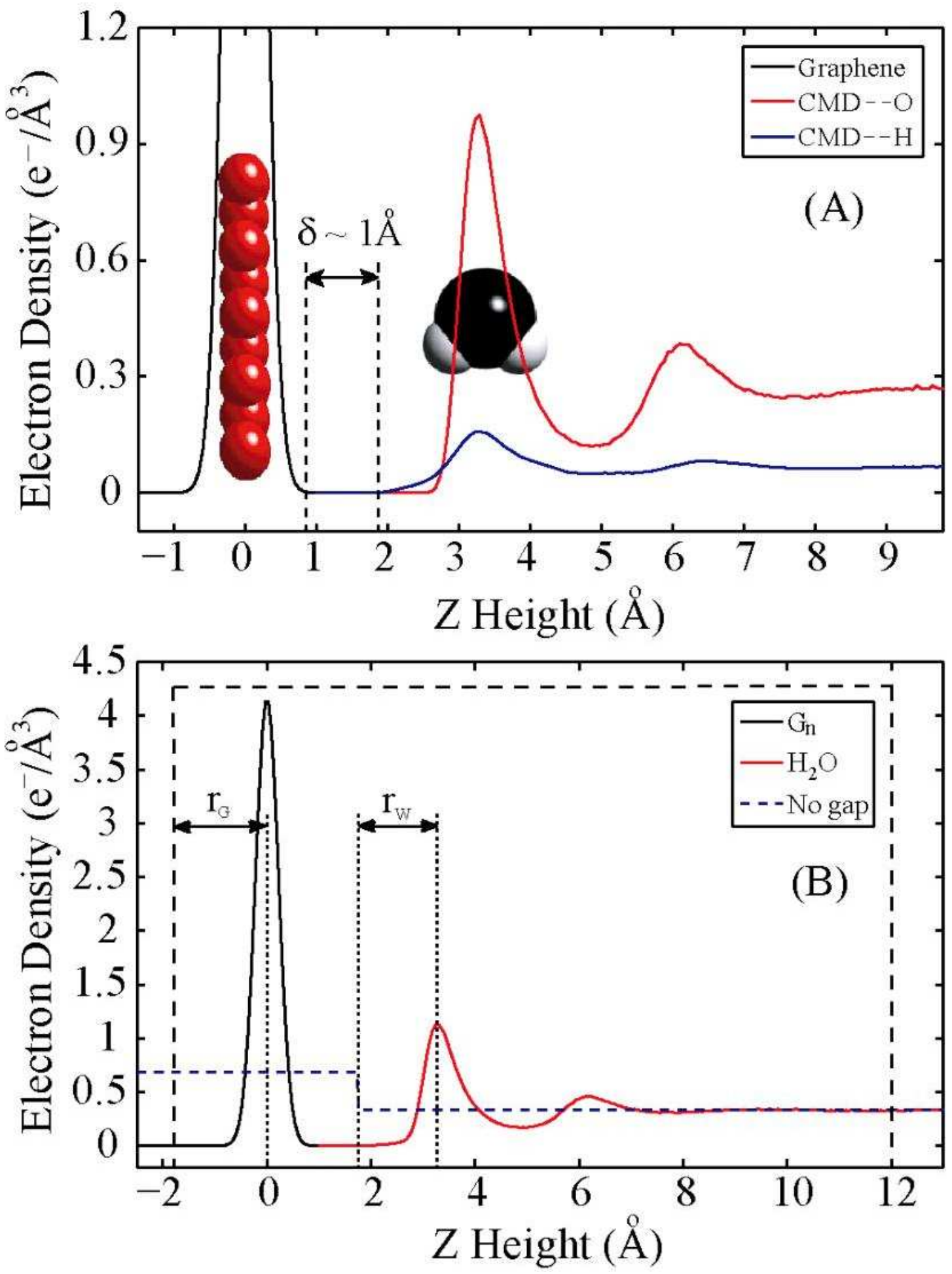